\documentclass{svjour3}                     
\smartqed  
\usepackage{graphicx}
\usepackage{mathptmx}      
\usepackage{natbib}
\usepackage{amssymb}
\journalname{SSRv}
\newcommand{\cc}{cm$^{-3}$}

\newcommand{\aap}{Astron.\ Astrophys.\ }
\newcommand{\apj}{Astrophys.\ J.\ }
\newcommand{\apjl}{Astrophys.\ J.\ Letters }
\newcommand{\nat}{Nature }
\newcommand{\jgr}{J.\ Geophys.\ Res. }
\begin{document}

\title{The Origins and Physical Properties of the Complex of Local
Interstellar Clouds}

\titlerunning{Origins and Physical Properties of the CLIC}

\author{Jonathan D. Slavin} 

\institute{Harvard-Smithsonian Center for Astrophysics, 60 Garden Street, MS
83, Cambridge, MA 02138, USA\\
Tel.: +161-74-967981\\
\email{jslavin@cfa.harvard.edu}
}

\date{Received: date / Accepted: date}

\maketitle

\begin{abstract}
The Complex of Local Interstellar Clouds (CLIC) is a relatively tight grouping
of low density, warm, partially ionized clouds within about 15 pc of the Solar
System.  The Local Interstellar Cloud (LIC) is the cloud observed on most
lines of sight and may be the cloud that immediately surrounds our Solar
System, the properties of which set the outer boundary conditions of the
heliosphere.  Using absorption line data toward nearby stars, \emph{in situ}
observations of inflowing interstellar gas from spacecraft in the Solar
System, and theoretical modeling of the interstellar radiation field and
radiative transfer, we can deduce many characteristics of the LIC.  We find
that the LIC is partially ionized with modest electron density, $n_e \approx
0.07$ \cc.  The combination of its temperature and ionization favor
photoionization/thermal equilibrium over a non-equilibrium cooling cloud
picture.  The abundances in the LIC suggest moderate dust destruction for
silicate dust but complete destruction of carbonaceous grains.  An origin for
the LIC as a density enhancement in the ambient medium that has been overrun
by a shock seems likely, while its velocity away from the Sco-Cen association
points to a possible connection to that region and the Loop I bubble.
\end{abstract}

\section{Introduction}
\label{intro}
The interstellar medium (ISM) that surrounds the Solar System impinges on the
outflowing Solar wind creating the heliosphere.  The particular
characteristics of that circumheliospheric interstellar medium (CHISM), such
as its density, ionization, temperature and magnetic field, determine in
detail the interaction of the gas with the solar wind.  Thus understanding the
nature of the heliosphere requires knowledge of the state of the CHISM while
at the same time interpretation of some of the \emph{in situ} observations
relevant to discerning the nature of the CHISM demands an understanding of the
heliosphere.

It has long been assumed, because it appears on so many lines of sight, that
the velocity component (or cloud) identified as the Local Interstellar Cloud
(LIC) surrounds the heliosphere.  Recently \citet{Redfield+Linsky_2008} have
analyzed many lines of sight toward nearby stars and identified 15 distinct
velocity vectors, which are attributed to coherent clouds in the local ISM.
The LIC component, found in by far the most lines of sight, was found to have
an associated temperature of $T = 7500 \pm 1300$ K.
\citet{Redfield+Linsky_2008} claim that the heliosphere is located in a
transition zone between the LIC and G cloud based on the fact that the
temperature and velocity of the gas are intermediate between those found for
the LIC and G cloud components.  We believe that this is somewhat speculative
as yet since the temperature of the inflowing gas is within the error bars on
the LIC while falling outside of the error bars for the G cloud value.  The
inflowing gas velocity does appear to be somewhat discrepant ($2-3$ km
s$^{-1}$) with their LIC vector, but that may be due to some weak disturbance
in the cloud.  Much of what we discuss does not depend specifically
on whether the heliosphere is within the LIC as long as their properties are
not very different, which appears to be the case.  The primary assumption
that we make related to this is that the portion of the line of sight between
the Solar System and the star $\epsilon$~CMa that traverses the LIC can be
treated as passing through a cloud that is in thermal and photoionization
equilibrium.

\section{The Physical Characteristics of the Complex of Local
Interstellar Clouds}
\subsection{Primary Observational Evidence}
The complex of local interstellar clouds, of which the Local Interstellar
Cloud (LIC) that surrounds the Solar System is one member, are nearby, low
density and partially ionized patches of gas that are surrounded by much lower
density and (probably) hot ISM, the Local Bubble.  We have the most
information about the LIC since most sight lines to nearby stars pass through
it, but the LIC is not extremely different from other clouds
within the CLIC \citep[see][also Redfield, this volume]{Redfield+Linsky_2008}.

Among the information on the LIC that we need to understand its origins are:
\begin{itemize}
\item ionization -- is the LIC gas in photoionization equilibrium or not?,
\item abundances -- important for evaluating ionization corrections, radiative
cooling rate and as evidence for dust destruction or enrichment,
\item temperature/density/pressure -- where does the LIC fit into typical ISM
phase picture? Is the LIC in thermal equilibrium?,
\item velocity -- can provide hints as to the dynamics, history and past
environment of the LIC,
\item magnetic field -- both the strength and orientation affect the size and
shape of the heliosphere.  The field also provides pressure support for the
cloud and can play an important role in determining the nature of the cloud
boundary.
\end{itemize}
Some of this information is relatively directly measureable, e.g.\ the
temperature of the cloud at the Solar System via He$^0$ temperature
observations \citep{Witte_2004}, but most require some modeling to infer their
values.

Evidence on the nature of the LIC comes from a wide variety of sources.  There
is a large database of ion absorption lines toward nearby stars wherein
particular velocity components have been identified as being due to the LIC
\citep[see][and references therein]{Redfield+Linsky_2008}.  This has been
accomplished by finding a consistent velocity vector for the cloud given the
observed projections for many lines of sight.

Among the important ions for which column densities have been observed for the
LIC are: Mg\,\textsc{ii}, Mg\,\textsc{i}, C\,\textsc{ii}$^*$, S\,\textsc{ii},
and Fe\,\textsc{ii}.  The importance of these ions lies in the constraints
they place on the physical state of the cloud.  As we discuss below, with
Mg\,\textsc{ii} and Mg\,\textsc{i} we can determine the electron density.
With the addition of C\,\textsc{ii}$^*$, Si\,\textsc{ii} and Fe\,\textsc{ii} we
gain information on the elemental abundances of dust components. These along
with  S\,\textsc{ii} provide necessary information on the total cooling within
the cloud, which is primarily due to forbidden line emission.  In order to
understand the ionization and thermal properties of the cloud it is necessary
to have data on these column densities and data on more ions will generally
help to constrain the models better.

The most direct evidence on the state of the CHISM comes from observations of
neutral He.  Because of its small cross section for charge transfer reactions,
He$^0$ is believed to sail through the bowshock, heliopause and termination
shock essentially unaffected.  As a result the measurements of He$^0$ density
and temperature, $n(\mathrm{He}^0) = 0.015 \pm 0.003$ \cc\ and
$T(\mathrm{He}^0) = 6300 \pm 340$ K \citep{Witte_2004}, are the best
constraints we have on the CHISM.  Other \emph{in situ} data including
backscattered H Lyman $\alpha$, anomalous cosmic rays and pickup ions provide
further constraints, but each comes with a more model-dependent
interpretation.  Direct observations of interstellar dust flowing into the
Solar System \citep{Baguhl_etal_1995,Landgraf_etal_2000} is very interesting
for what it tells us about the chemical composition of the LIC and clues about
its history, but does not directly inform us about the gas phase abundances
that govern its thermal balance.  Dust can be an important heat source in the
ISM via photoelectron ejection, but we find the dust heating is small relative
to photoionization heating for the conditions of the LIC.

An additional set of input data needed to understand the physical conditions
in the LIC is the ionizing radiation field. Both the current ionization and
the sources of photoionization are needed if we are to make sense of the
present state of the cloud. Portions of the field have been directly observed,
namely the far UV and extreme UV from stellar sources.  There have also been
direct observations of diffuse soft X-rays, though the source of those remain
controversial (see Koutroumpa, this volume).   As of now it appears that most
of the softest X-rays (Wisconsin B and C band, $E = 70 - 280$ eV) do come from
hot gas within the Local Bubble and these are the most important X-rays for
the ionization of the cloud.  Unfortunately the potentially even more
important diffuse EUV background has yet to be observed and instrumental
limitations make it unlikely that such an observation can be made for some
time to come.

Another important observational fact is that the LIC, and indeed the CLIC,
exist in a large, extremely low density cavity.  The location of these clouds
within the Local Bubble is clearly important to understanding their origins
and evolution.

\begin{figure}[ht!]
\includegraphics[width=4.5in]{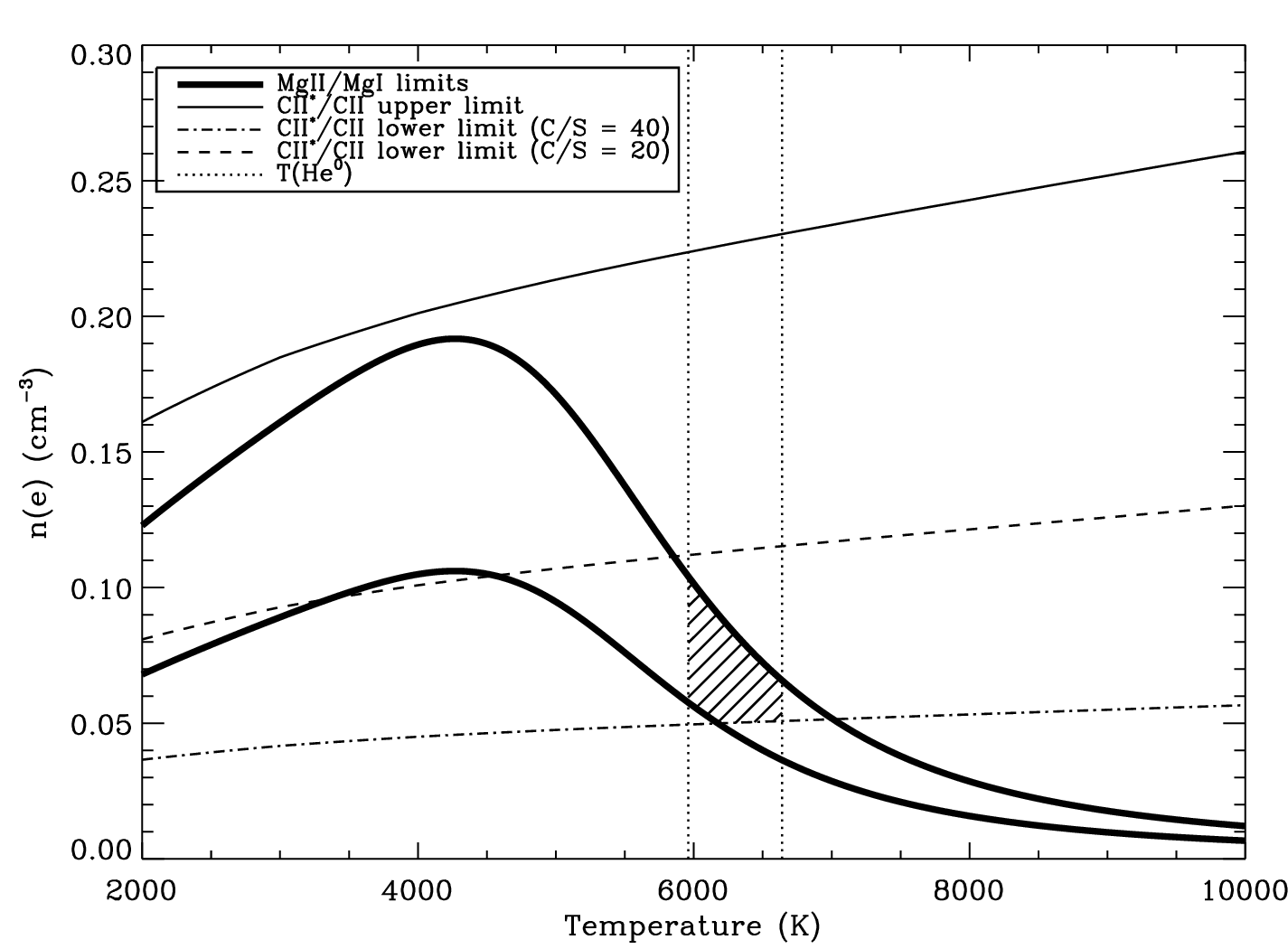}
\caption{Constraints on the electron density and temperature in the LIC
derived from Mg\,\textsc{ii}, Mg\,\textsc{i}, C\,\textsc{ii}$^*$, 
C\,\textsc{ii} and \emph{in situ} He$^0$ temperature observations.}
\label{fig:ne_T}
\end{figure}

\subsection{Electron Density and Temperature of the LIC}
As noted above, the observations of Mg\,\textsc{ii} and Mg\,\textsc{i} place
constraints on the electron density of the LIC.  Balance of FUV ionization,
radiative and dielectronic recombination and charge transfer gives us
\begin{equation}
n_e = \frac{\Gamma(\mathrm{Mg}^0)}{\alpha(\mathrm{Mg}^+) +
C^{CT}(n(\mathrm{H}^+)/n_e)}\frac{N(\mathrm{Mg\,I})}{N(\mathrm{Mg\,II})}
\end{equation}
where $\Gamma($Mg$^0)$ is photoionization rate, $\alpha(\mathrm{Mg}^+)$ is
(total) recombination rate and $C^{CT}$ is charge transfer rate. We derive the
photoionization rate using the observed and modeled FUV background from
\citet{Gondhalekar_etal_1980}.  If we have $N(\mathrm{C\,II})$ and
$N(\mathrm{C\,II}^*)$ we can also get $T$:
\begin{equation}
\frac{N(\mathrm{C\,II}^*)}{N(\mathrm{C\,II})} = \frac{\gamma_{12}(T)\,n_e}
{A_{21} + \gamma_{21}(T)\,n_e}
\end{equation}
Unfortunately, the most easily observable C\,\textsc{ii} line at 1334.5\AA\ is
nearly always saturated, making the column density difficult to derive with
any certainty.

For the $\epsilon$ CMa line of sight, $n_e$ and $T$ have been derived by
\citet{Gry+Jenkins_2001} by making assumptions for the abundance of C to S and
using the well observed S\,\textsc{ii} line results to derive a upper limit
for $N($C\,\textsc{ii}). If we use the constraints on $T$ from \emph{in situ}
observations and up-to-date values for the Mg$^+$ recombination coefficients, 
we find that we need an abundance ratio of C/S $> 20$ to find a viable
solution for $n_e$ and $T$.  The combination of the upper limit on 
$N($C\,\textsc{ii}), the observed limits on Mg\,\textsc{ii}/Mg\,\textsc{i} and
the observed limits on $T$ lead to tight limits, $n_e = 0.05 - 0.104$ \cc, as
illustrated in Figure \ref{fig:ne_T}. We discuss these limits and the
implications for the C abundance in more detail in \citet{Slavin+Frisch_2006}.
When we carry out more detailed modeling including models for the interstellar
radiation field, thermal balance and radiative transfer
\citep{Slavin+Frisch_2008}, we derive even tighter limits on $n_e$, $n_e =
0.07 \pm 0.01$ \cc.

\subsection{Radiative Transfer Models}
To go from ion column densities to abundances requires ionization corrections,
especially for $N($H). In general elements with first ionization potential
$E_0 > 13.6$ eV will require an ionization correction to derive the total
element column density and these include He, N, O, Ne, and Ar. Oxygen is a
special case, however, because its ionization is tightly coupled to H
ionization by charge transfer. Nitrogen is also similarly coupled to H
ionization but much less tightly.  These corrections are particularly
important because the H\,\textsc{i} column is not very well determined in many
cases (e.g. $\epsilon$ CMa).  Data from the Extreme Ultraviolet Explorer
(EUVE) can give us the total H\,\textsc{i} column, but does not give us the
fraction of the total attributable to each velocity component if there are
multiple velocity components along the line of sight.

In order to derive the ionization corrections one must have a model for the
ionization of the cloud and in general this demands a model for the ionizing
radiation field, as well as the radiative transfer within the cloud.  To model
the field we use the sum of emission from stellar sources (FUV and EUV) and
diffuse sources including the hot gas that we assume fills the Local Bubble
and an evaporative boundary between the warm cloud and the hot gas. 

The nature of the boundary between the warm cloud gas and the surrounding hot
gas is still quite uncertain as there have been no definitive observations
establishing the existence of an evaporative boundary.  It is known that the
magnetic field will suppress thermal conduction across field lines, yet a
completely magnetically isolated cloud would seem unlikely.  In our models we
assume a magnetic suppression of conductivity of a factor of 2, appropriate to
the case in which the field is at a 45$^\circ$ angle with the cloud surface.
This part of the radiation field is clearly quite uncertain and we intend to
explore possible variants in future work.

\begin{figure}[ht!]
\includegraphics[width=4.5in]{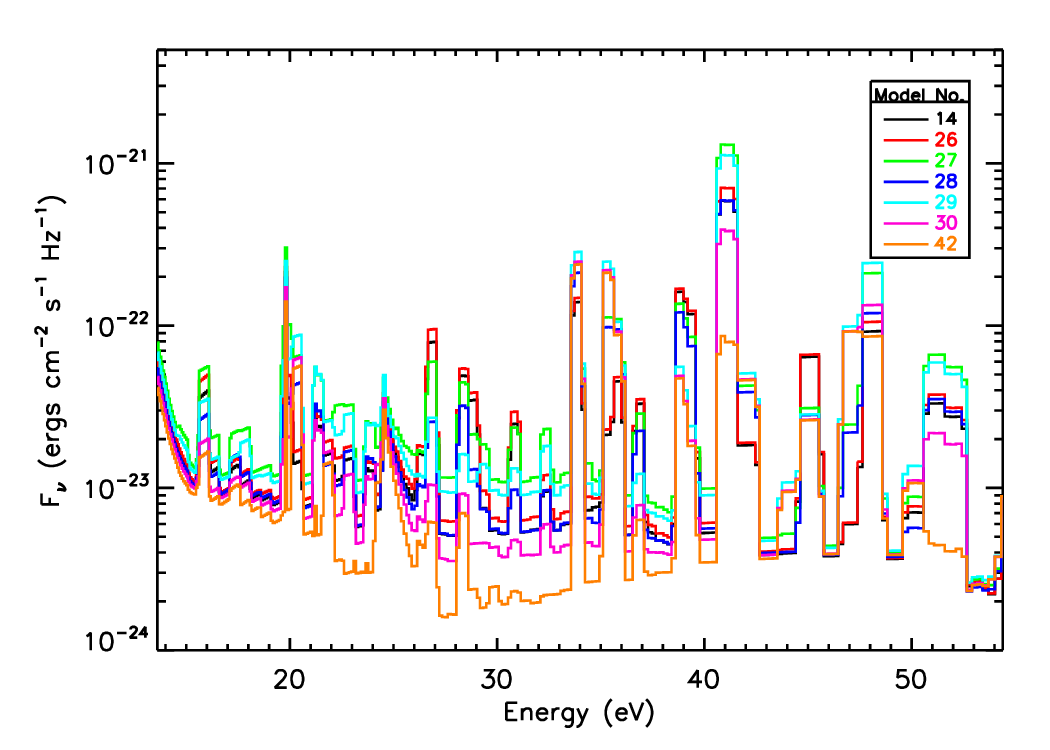}
\caption{Model ionizing radiation fields for several different models from
\citet{Slavin+Frisch_2008}, all of which match the observational constraints.}
\label{fig:radfield}
\end{figure}

We find that the details of the field do not seem to strongly influence our
results for the ionization of the cloud. We show in Figure \ref{fig:radfield}
the H ionizing radiation field at the position of the Sun in several models
that are consistent with the observations. The requirements we impose on the
models to match $n($He$^0)$, $T($He$^0)$ and the ion column densities act to
fix the H ionization, $X(\mathrm{H})$ and $n_e$.

We find successful models for range of values for $N(\mathrm{H\,I})$ towards
$\epsilon$CMa and $T$(hot gas). Our results for the Solar location are:
\begin{itemize}
\item $n(\mathrm{H}^0) = 0.19 - 0.20$ cm$^{-3}$,
\item $n_e = 0.07 \pm 0.01$ cm$^{-3}$,
\item $n(\mathrm{H}) = 0.23 - 0.27$ cm$^{-3}$,
\item $B = 2.1, 2.5 \mu$G for two best models (the magnetic field affects
emission intensity from cloud boundary).
\end{itemize}

\section{Dust and Elemental Abundances in the LIC}
\begin{figure}[ht!]
\includegraphics[width=4.5in]{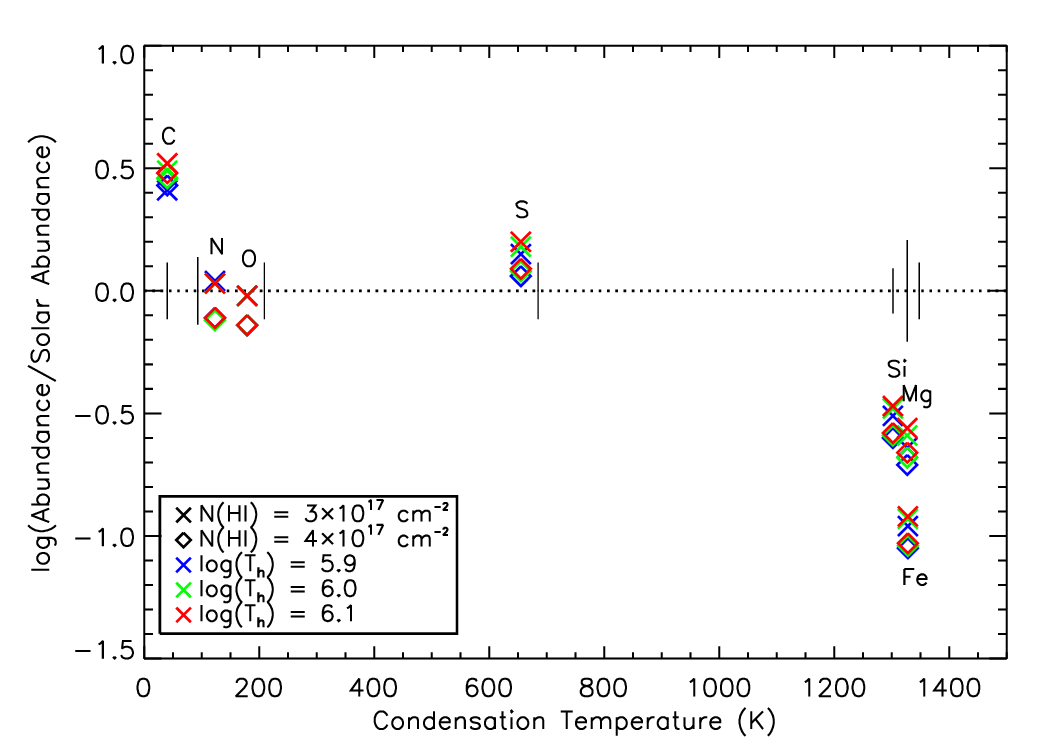}
\caption{Abundances relative to \citet{Asplund_etal_2005} solar abundances vs.
condensation temperature. Results from models of the ionization for the
$\epsilon$ CMa line of sight including radiative transfer
\citep{Slavin+Frisch_2008}.  The symbol shape indicates the assumed
H\,\textsc{i} column density for the model, while the color indicates the
assumed temperature for the hot gas of the Local Bubble.  The abundances are
fixed so as to match the observed ion column densities.}
\label{fig:abund}
\end{figure}

Our approach to modeling the abundances in the LIC is to force our models to
match the column densities observed by adjusting the elemental abundances.
Thus the abundances are an output of the modeling rather than in input.
Figure \ref{fig:abund} illustrates the abundance results for C, N, O, S, Si,
Mg and Fe relative to a particular assumed solar abundance of the elements
\citep{Asplund_etal_2005}.  The x-axis is the condensation temperature for the
element, a quantity often presumed to correlate with the amount of depletion
from the gas phase. In Table \ref{tab:abund} we list the derived abundances in
our models that are consistent with observations.

\begin{table}
\caption{Elemental Gas Phase Abundances (ppm)}
\label{tab:abund}
\begin{tabular}{crrrrrrr}
\hline\noalign{\smallskip}
& \multicolumn{7}{c}{Element} \\ 
\cline{2-8} \\
Model No. & C & N & O & Mg & Si & S & Fe \\
\noalign{\smallskip}\hline\noalign{\smallskip}
14 & 589 &  40.7 & 295 & 5.89 &  7.24 &  14.1 & 2.24 \\
25 & 631 &  66.1 & 437 & 7.76 &  10.0 &  19.5 & 3.09 \\
26 & 661 &  46.8 & 331 & 6.61 &  8.13 &  15.8 & 2.51 \\
27 & 759 &  64.6 & 437 & 8.71 &  10.7 &  20.9 & 3.31 \\
28 & 708 &  45.7 & 331 & 7.08 &  8.32 &  16.6 & 2.57 \\
29 & 813 &  64.6 & 437 & 9.33 &  11.0 &  21.9 & 3.39 \\
30 & 741 &  46.8 & 331 & 7.41 &  8.51 &  17.0 & 2.63 \\
42 & 724 &  39.8 & 295 & 6.76 &  7.76 &  15.1 & 2.34 \\
\noalign{\smallskip}\hline
\end{tabular}
\end{table}

\begin{figure}[ht!]
\includegraphics[width=4.5in]{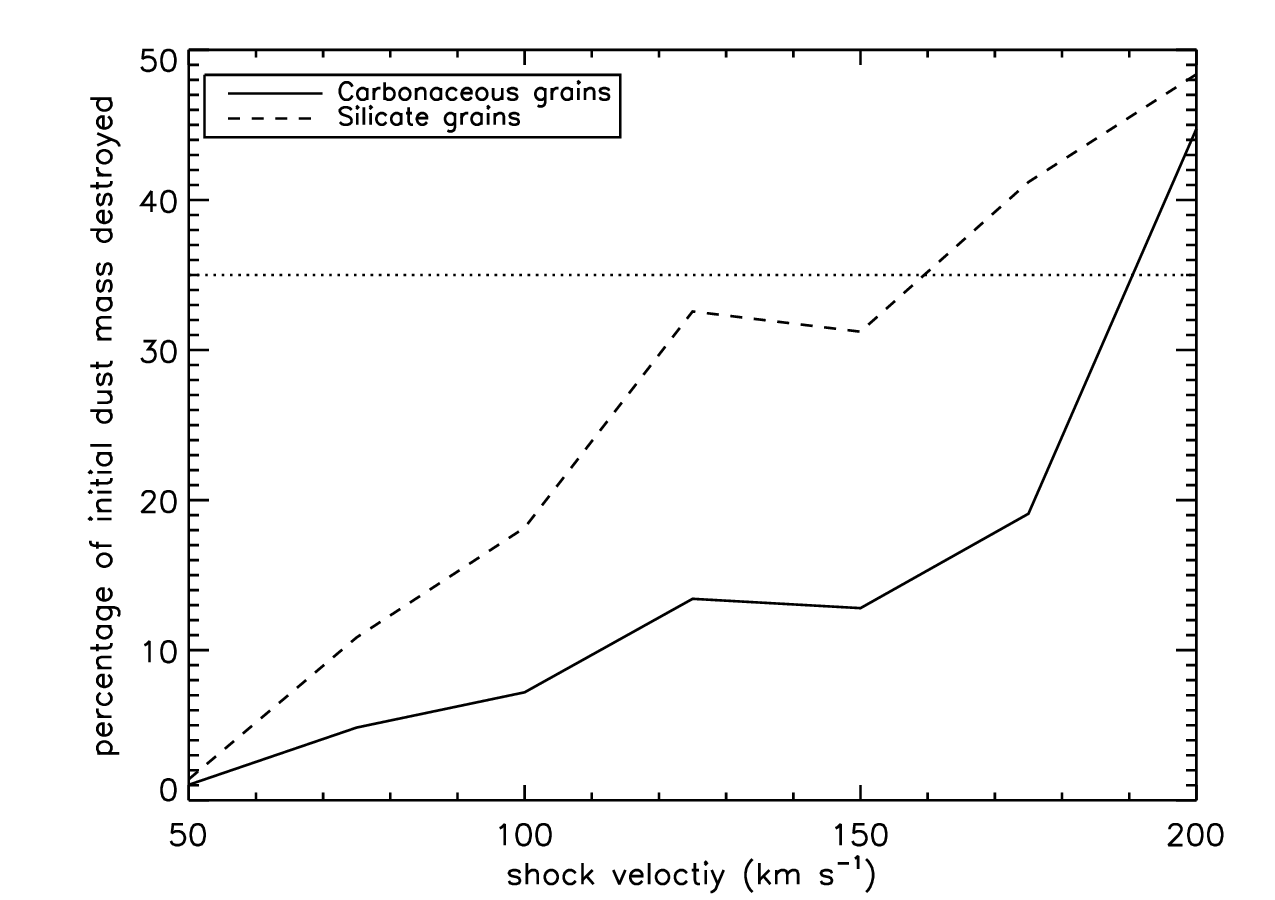}
\caption{Percentage of initial total grain mass lost because of dust
destruction in the shock vs.\ shock speed.  To get the Si gas phase abundance
determined for the LIC we need $\lesssim$35\% destruction of silicate dust
which is consistent with $v_\mathrm{shock} \lesssim 150$ km s$^{-1}$.  Such
shocks should destroy less than 15\% of the cabonaceous dust, however, in
contradiction with the derived large gas phase abundance of C.}
\label{fig:dustdest}
\end{figure}

The derived abundances indicate modest depletion of the constituents of
silicate dust, Si, Fe, Mg, and O, implying that at least some destruction of
this type of dust.  Depending on ones assumptions about the initial depletion
of Si \citep[][quote values of 70-95\%]{Savage+Sembach_1996}, the total Solar
abundance of Si (recent determinations range from about 30 to 43 ppm) and its
current gas phase abundance (we find $\sim 7.2 - 11.5$ ppm) the fraction of
the silicate grain mass destroyed (i.e. returned to the gas phase) ranges from
0 to 35\%.  The high C abundance on the other hand seems to indicate that all
the carbonaceous dust has been destroyed.  Radiative shocks destroy dust via
various processes: sputtering, vaporization, shattering.  Detailed
calculations by \citet{Jones_etal_1996} find that silicate dust should be
\emph{more} destroyed than carbonaceous dust by shocks, as illustrated in
Figure \ref{fig:dustdest}.  Therefore either some other process is strongly
influencing the gas phase abundances in the LIC or the models of shock
processing of grains need revision.

\section{LIC Ionization and Thermal Balance}
\begin{figure}[ht!]
\includegraphics[width=4.5in]{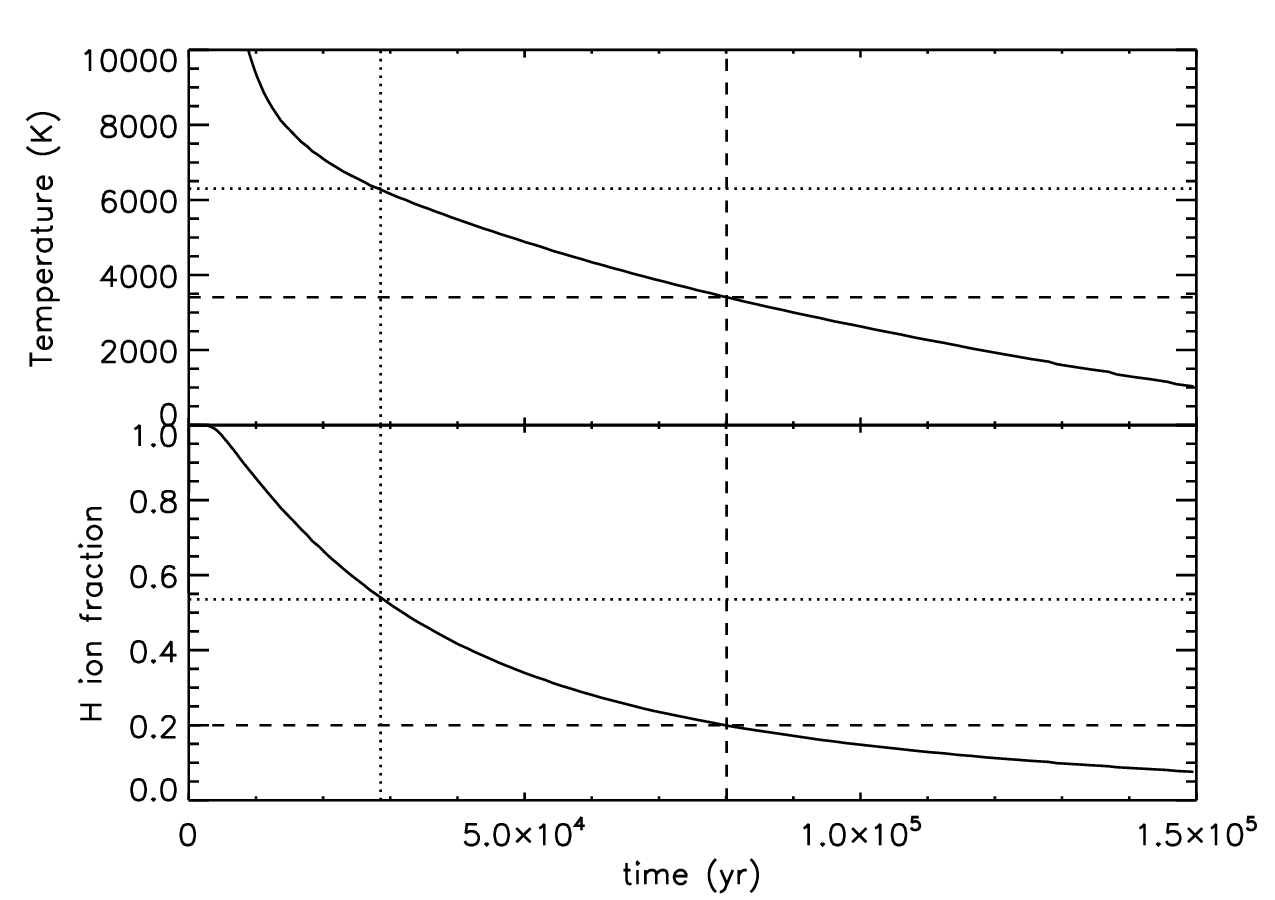}
\caption{Time evolution of temperature and H ion fraction behind a 100 km
s$^{-1}$ shock (J. Raymond, private communication).  The dotted lines show the
time, temperature and H ion fraction when the gas cools to the LIC
temperature, while the dashed lines show the same for the LIC H ion fraction.
Isobaric cooling shows a similar lag of H recombination.  The gas cools too
fast to allow the relatively low degree of ionization in the LIC at
temperatures of $\sim 6300$ K without some substantial heating source.}
\label{fig:ionevol}
\end{figure}

Observations with the Extreme Ultraviolet Explorer (EUVE) toward nearby stars
found unexpected results for the ratio of H\,\textsc{i} to He\,\textsc{i}
column density \citep{Dupuis_etal_1995}.  Instead of the expected ratio of
$\le 10$ that one would get if the cosmic He abundance is 0.1 and H is more
ionized than He, it was found that $N(\mathrm{H\,I})/N(\mathrm{He\,I}) \sim
14$, indicating that He is more ionized than H. This unusual ionization of the
local ISM has long been considered puzzling and has led to the suggestion that
the LIC is out of ionization equilibrium, being overionized for its temperature
because of an earlier ionizing event (e.g. a shock) \citep[see,
e.g.,][]{Lyu+Bruhweiler_1996}.  The long timescale for recombination,
particularly of H, it was reasoned, makes it likely that the LIC
is out of ionization equilibrium.  However, the cooling rate of the gas also
has to be considered in such a model.  Doing this one finds that in fact
the cooling time for the gas in any likely scenario is considerably less than
the recombination time.  As a result, if the LIC were cooling from a hotter
and more ionized state, it should still be quite highly ionized by the time
that it has cooled to the observed temperature of $T \approx 6300$ K. In
Figure \ref{fig:ionevol} we illustrate this by showing the temperature and
ionization evolution behind a 100 km s$^{-1}$ shock. Even greater disparity
between the cooling time and recombination time is found for a simple isobaric
cooling model.  Therefore the fact that the LIC is in fact mostly neutral,
$X(\mathrm{H}^+) \sim 0.2$, implies that the cloud has had time to recombine
while being maintained at a warm temperature.  This requires a heat source to
balance the cooling.  While alternative sources have been proposed, such as
turbulent dissipation \citep{Minter+Spangler_1997}, the most likely heat
source appears to be photoionization heating.  Such heating is also
accompanied by ionization, suggesting that the cloud is at least close to
thermal and photoionization equilibrium.

Another argument in favor of the cloud being in ionization equilibrium was
first suggested by \citet{Jenkins_etal_2000} based on Ar\,\textsc{i} and
O\,\textsc{i} data.  Since O ionization is tied to H ionization by charge
exchange, if we assume an O abundance we can then compare the ionization of Ar
and H.  In recombining gas, it is found that Ar and H have roughly equal
ionization fractions because H$^+$ and Ar$^+$ have similar recombination
coefficients.  However, for gas in photoionization equilibrium, Ar\,\textsc{i}
is deficient relative to H\,\textsc{i} (or O\textsc{i}) because the
photoionization cross section for Ar$^0$ is 5-30 times larger than that for
H$^0$.  Observations for the LIC \citep{Jenkins_etal_2000} find
$X(\mathrm{Ar}^0)/X(\mathrm{H}^0) \sim 0.4$ toward nearby white dwarfs.
Detailed NEI calculations for cooling gas show that 
$X(\mathrm{Ar}^0)/X(\mathrm{H}^0)$ remains $\sim 1$ until gas nears
equilibrium \emph{and} is photoionized.

Despite this evidence that the LIC is currently close to ionization and
thermal equilibrium, there are reasons to believe that it has not always been
so.  The LIC is clearly many times denser than its surrounding gas in the 
Local Bubble as can be deduced from the low absorption by neutral gas within
the bubble and lack of observable optical emission from possible warm ionized
gas that could conceivably fill the cavity. That leaves only highly ionized
and very low density gas as the primary volume filling gas in the bubble.
Therefore it appears highly likely that the gas that presently makes up the
LIC and other nearby clouds was at one time substantially overdense compared
with the surrounding medium before becoming incorporated into the Local
Bubble. The most likely scenario is that cold neutral medium gas, with $n \sim
100$ cm$^{-3}$, $T \sim 100$ K, that was embedded in warm gas was hit by a
shock. However, it is important to note that any shock \emph{no matter what
speed} hitting such a dense cloud will go radiative in the cloud.  Thus one
needs to find a means to heat the cloud to warm neutral medium temperatures.
An origin in a fragmented shell implies a similar radiative shock and heating
requirements.  The means to heat the shocked warm clouds seems to require
their expansion to lower density as the pressure of the bubble drops
at the same time as ionizing flux from the hot gas and possibly from the cloud
boundary regions provides heating.

\begin{figure}[h!]
\includegraphics[width=4.5in]{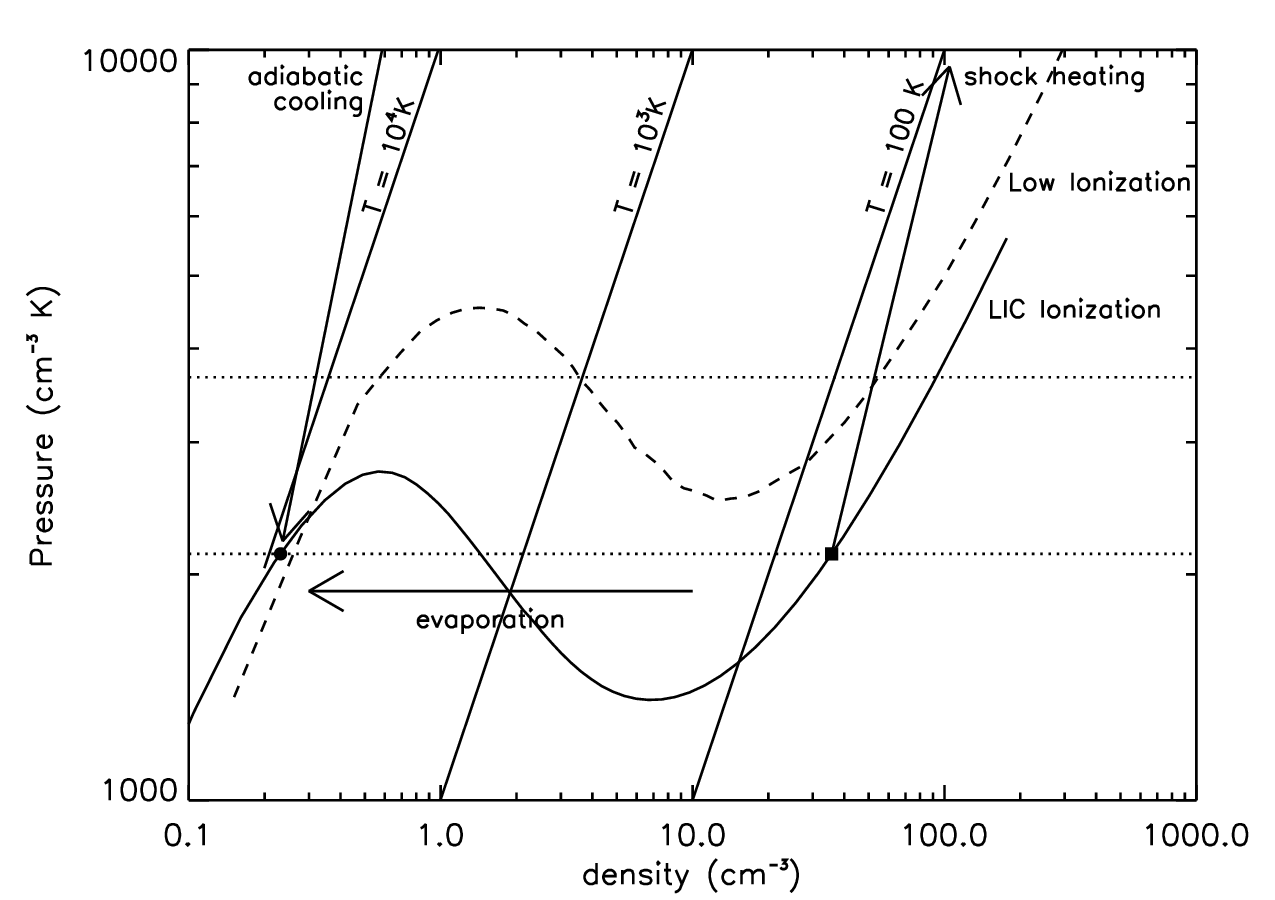}
\caption{Phase diagram for the diffuse interstellar medium for either low
ionizing flux conditions \citep[dashed curve, using rates
from][]{Wolfire_etal_2003} or the moderate ionizing flux as modeled for
the LIC (solid curve using our modeled radiation field).  Points on the curve
are conditions of thermal equilibrium, below the curve heating exceeds cooling
and above the curve cooling exceeds heating. The diagonal lines are curves of
constant temperature.  A shock will tend to move a gas parcel up and to the
right in the diagram, while adiabatic cooling moves points down and to the
left as illustrated by the arrows.  Evaporation via thermal conduction moves
points to left. Note that the pressure in the plot is only thermal pressure
and thus neglects the dynamically important magnetic, cosmic ray and turbulent
pressures in the ISM.}
\label{fig:phase}
\end{figure}

For diffuse ISM conditions, calculated heating and cooling rates typically
lead to the possibility of thermal balance with two stable thermal
phases within a limited range of thermal pressures, with a cold neutral phase
and a warm neutral or (perhaps partially) ionized phase. Figure
\ref{fig:phase} shows a density vs.\ pressure plot or phase diagram showing
two different phase equilibrium curves.  The one for ``Low Ionization'' comes
from the work of \citet{Wolfire_etal_2003} and assumes low ionizing flux
whereas the ``LIC ionization'' one is calculated using one of our model
ionizing radiation fields for the LIC. We note that the thermal pressure will
generally not dominate the total dynamical pressure because other pressure
forms including magnetic, cosmic ray and turbulent, are typically estimated to
be of the same order of magnitude as the thermal pressure.  This does not
affect the phase curves, however, since it is the components of the thermal
pressure (i.e. density and temperature) that directly affect the
heating-cooling balance.  In the diffuse ISM cosmic ray heating is small
compared to dust and photoionization heating.  It may be that turbulent
dissipation, particularly in concert with MHD turbulence, provides significant
heating, however the rate for that remains quite uncertain and is neglected in
the Figure.

The arrows on the plot indicate how gas parcels will evolve under the
influence of shocks, adiabatic cooling (cooling via expansion) and evaporation
via thermal conduction.  The shock arrow indicates a relatively small increase
in pressure, which would require only a mach 2.5 (relative to the cold gas)
shock.  A shock that could heat typical warm (ionized or neutral) medium gas
to about $10^6$ K would need to be much faster, $v_s \approx 270$ km s$^{-1}$,
or mach 27 in the warm medium.  The pressure would thus be increased to $P/k_B
\sim 2\times10^6$ cm$^{-3}$ K.  In order for the local clouds to become warm
would require the pressure to drop by more than two orders of magnitude after
the shock passed over them.  This could be achieved after sufficient
expansion, e.g.\ a factor of $\sim 4$ in radius assuming adiabatic expansion
of a spherical bubble.  This requires that the clouds were close enough to the
center of the superbubble that the shock had not gone radiative yet and that
the clouds (or at least a fraction of them) could survive long enough to
persist until our current state in which the surrounding bubble has a
relatively low pressure.
 
\section{The Origin the Complex of Local Interstellar Clouds}
The above discussion lays out some of the challenges facing any model for the
origin of the complex of local interstellar clouds.  In summary we would like
a theory to explain these facts: \begin{itemize}
\item The density, temperature and ionization of the clouds are in sharp
contrast to the surrounding Local Bubble gas (though we don't know all the
properties of that gas),
\item the CLIC has a significant velocity relative to the LSR and direction
roughly away from Galactic center,
\item the ionization of the LIC is unusual with He apparently more ionized
than H,
\item the abundances in the gas seem to imply that carbonaceous dust has been
destroyed, and yet interstellar dust observed in the Solar System implies a
relatively low gas-to-dust ratio.
\end{itemize}

A number of theories have been put forward to explain the CLIC. The clouds
have been variously proposed to be: 1) pieces of the Sco-Cen bubble from an
earlier epoch of star formation \citep{Frisch_1981}, 2) a fragment from
Sco-Cen/LB interaction \citep{Breitschwerdt_etal_2000}, and 3) a flux
tube/filament that has broken away from the bubble wall
\citep{Cox+Helenius_2003}.  We would add to this list, 4) a dense cloud in the
ambient medium overrun by an expanding bubble shock, a model that we have
discussed briefly above but that has yet to be fully explored.  

Each of these models has its problems.  The velocity and relative positions
of the CLIC and Loop I bubble strongly suggest a connection between them but
detailed modeling of how these clouds could have come from that bubble is
lacking.  \citet{Breitschwerdt_etal_2000} propose that the Local Bubble and
the Loop I bubble are interacting and that the CLIC is associated with the
wall that separates the bubbles. In their model the clouds are created 
by instabilities generated in the interaction region.  The LIC is currently
about 70 pc from that neutral wall and moving about 20 km s$^{-1}$ away from
the Sco-Cen association that is believed to be responsible for creating the
Loop I bubble.  It is unclear how cloudlets like the CLIC could have been
traveling for 3.5 million years away from this interaction zone and yet the
wall is apparently intact between the two bubbles.  \citet{Frisch_1981}
suggests that the clouds as well as the Local Bubble are associated with a
previous epoch of star formation of Sco-Cen. This requires that somehow a cold
neutral wall was reformed within the bubble between these epochs of star
formation.  The mechanism for doing that is left unexplained.  The flux tube
theory of the origins of the CLIC by \citet{Cox+Helenius_2003} requires that a
flux tube sprang from the wall of the Local Bubble pulling warm gas along with
it into the bubble interior.  The magnetohydrodynamics of this explanation
seem questionable however, in particular that one flux tube can spring from
the bubble wall while the rest of the bubble is not collapsing. Finally, our
idea that the clouds originated as cold clouds in a warm intercloud medium
seems reasonable but does not explain why the velocity of the CLIC is directed
away from the Sco-Cen association and towards the center of the Local Bubble
rather than away from it. We must appeal to a random velocity of the gas prior
to being overrun by the expanding Local Bubble to explain this.

\section{Summary}
The wide range of data that we have on the LIC has lead to a fairly complete
picture of the cloud.  We find that it is:
\begin{itemize}
\item  partially ionized, $X(\mathrm{H}^+) \sim 0.2 - 0.3$, $X(\mathrm{He}^+)
\sim 0.3 - 0.4$, $n_e \approx 0.07$ cm$^{-3}$ 
\item  has experienced mixed dust destruction -- moderate for silicate dust,
complete for carbonaceous,
\item  at or close to ionization equilibrium
\end{itemize}
An origin as a cloud embedded in a lower density medium that was shocked
seems likely, and some association with the Loop I bubble and Sco-Cen OB
association remains a possibility. Many mysteries remain about its abundances
and origins within the local ISM.

\begin{acknowledgements}
I would like to thank the organizers of the ``Outer Heliosphere to the Local
Bubble'' conference for inviting me to give this talk and Priscilla Frisch, my
collaborator in much of the work I presented.  This research was supported by
NASA Solar and Heliospheric Physics Program grants NNG05GD36G and NNG06GE33G
to the University of Chicago.
\end{acknowledgements}

\newcommand{\bibfont}{\footnotesize}

\end{document}